\documentclass{PoS}

\title{Interpreting the 750 GeV diphoton signal as technipion}

\ShortTitle{Interpreting the 750 GeV diphoton signal as technipion} 

\author{Piotr LEBIEDOWICZ\\
Institute of Nuclear Physics, Polish Academy of Sciences, PL-31-342 Cracow, Poland\\
E-mail: \email{Piotr.Lebiedowicz@ifj.edu.pl}} 
\author{Marta {\L}USZCZAK\\
Department of Theoretical Physics, University of Rzesz{\'o}w, PL-35-959 Rzesz{\'o}w, Poland\\
E-mail: \email{luszczak@univ.rzeszow.pl}}

\author{Roman PASECHNIK\\
Department of Astronomy and Theoretical Physics, Lund University, 
SE-223 62 Lund, Sweden
E-mail: \email{Roman.Pasechnik@thep.lu.se}}

\author{\speaker{Antoni SZCZUREK}\\
Institute of Nuclear Physics, Polish Academy of Sciences, PL-31-342 Cracow, Poland
\footnote{Also at University of Rzesz\'ow, PL-35-959 Rzesz{\'o}w, Poland.}\\
E-mail: \email{Antoni.Szczurek@ifj.edu.pl}}

\abstract{

We discuss whether the enhancement in the diphoton final state at 
$M_{\gamma \gamma}$ = 750 GeV, 
observed recently by the ATLAS and CMS Collaborations could be 
a neutral pseudoscalar technipion $\tilde{\pi}^0$. 
We considered two distinct minimal models
for the dynamical electroweak symmetry breaking. 
Here we concentrate only on two-flavor vector-like technicolor model and we assume that 
the two-photon fusion is a dominant production mechanism. 
We include contributions of $2 \to 1$, $2 \to 2$ and $2 \to 3$ partonic processes.
All the mechanisms give similar contributions to the cross section. 
With the strong Yukawa (technipion-techniquark) 
coupling $g_{TC} \simeq 20$ we roughly obtain the measured cross section of the ``signal''. 
With such value of $g_{TC}$ we get a relatively small total decay width $\Gamma_{\rm tot}$. 
We discuss also the size of the signal at lower energies (LHC, Tevatron) 
for $\gamma \gamma$ final states, where the enhancement was not observed.
We predict a measurable cross section for neutral technipion 
production associated with one or two soft jets. 
The technipion signal is compared with the Standard Model diphoton background contributions. 
We observe the dominance of inelastic-inelastic $\gamma \gamma$ processes. 
We predict the signal cross section for purely exclusive $p p \to p p \gamma \gamma$ procesess 
at $\sqrt{s}$ = 13 TeV to be about 0.2 fb. Such a cross section 
would be, however, difficult to measure with the planned integrated 
luminosity. 
We conclude that in all considered cases the signal is below the background or/and below 
the threshold set by statistics although some tension can be seen.
}

\FullConference{XXIV International Workshop on Deep-Inelastic Scattering and Related Subjects\\
          11-15 April, 2016\\
          DESY Hamburg, Germany} 
          
\begin{document}

\section{Introduction}

Recently both the ATLAS and CMS Collaborations announced an observation 
of an enhancement in the diphoton invariant mass at 
$M_{\gamma \gamma}\approx750$~GeV in proton-proton collisions at $\sqrt{s} =13$~TeV 
\cite{ATLAS:2015,CMS:2015dxe,Aaboud:2016tru,Khachatryan:2016hje}.
Remarkably, such a hint to a possible New Physics signal has triggered 
a lot of activities recently. 
Several possible interpretations were discussed (see for instance \cite{Strumia:2016wys}).

One of the appealing and consistent classes of technicolor (TC) models with 
a vector-like (Dirac) UV completion is known as the vector-like TC (VTC)
scenario~\cite{Kilic:2009mi}.
The simplest version of the VTC scenario applied to the EWSB possessed 
two Dirac techniflavors and a SM-like Higgs 
boson~\cite{Pasechnik:2013bxa,Lebiedowicz:2013fta,Pasechnik:2014ida}. 
Recently, the concept of Dirac UV completion has also emerged 
in composite Higgs boson scenarios with confined $SU(2)_{\rm TC}$ symmetry 
\cite{Cacciapaglia:2014uja,Hietanen:2014xca}.

\begin{figure}[!ht]
\centering
\includegraphics[width=0.17\textwidth]{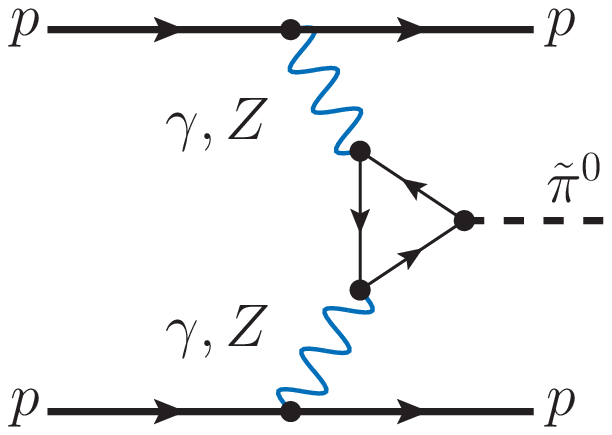}
\includegraphics[width=0.17\textwidth]{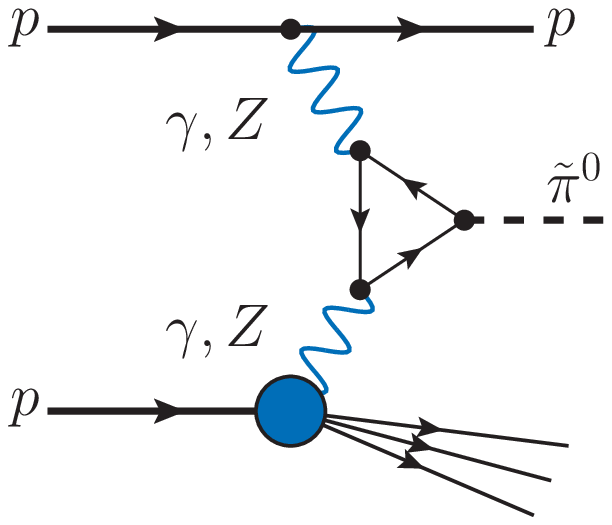}
\includegraphics[width=0.17\textwidth]{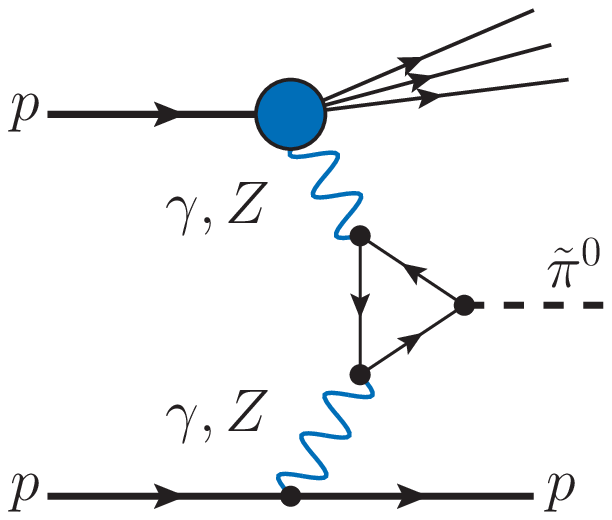}
\includegraphics[width=0.17\textwidth]{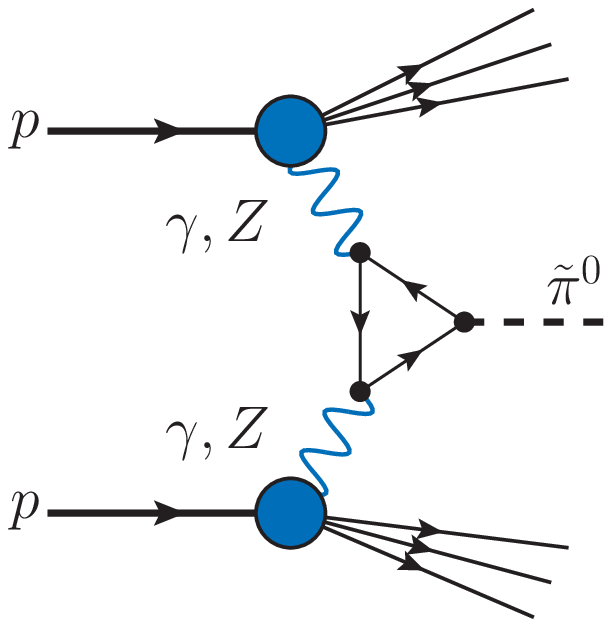}
  \caption{\label{fig:diagrams_order1}
  \small
Diagrams of neutral technipion production 
via the $\gamma \gamma$, $\gamma Z$ and $ZZ$ fusion 
in $pp$-collisions.}
\end{figure}

\begin{figure}[!ht]
\centering
\includegraphics[width=0.2\textwidth]{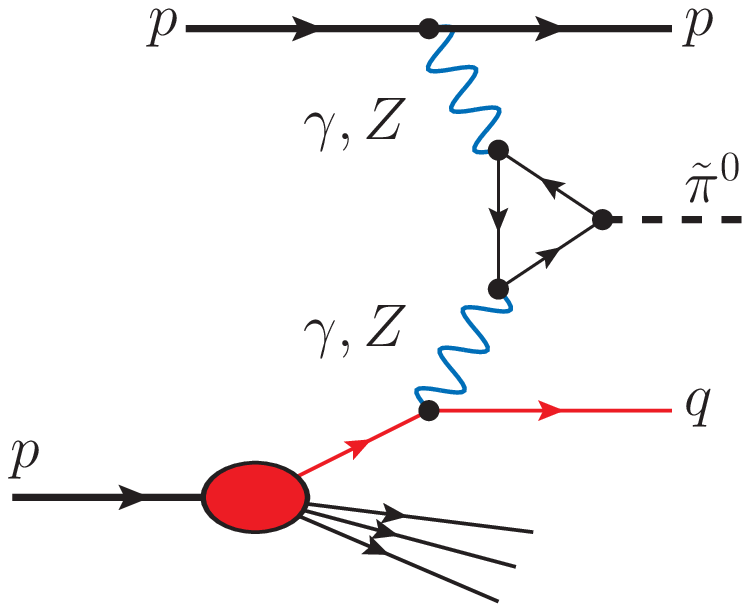}
\includegraphics[width=0.2\textwidth]{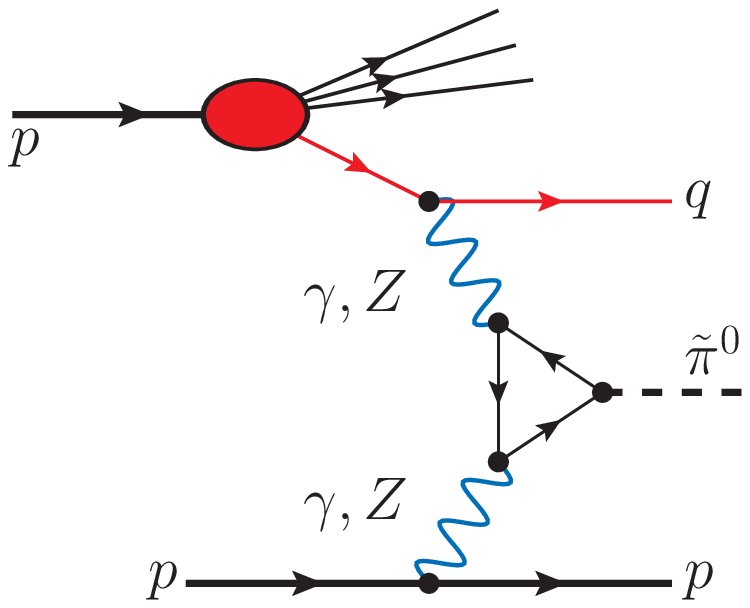}
\includegraphics[width=0.2\textwidth]{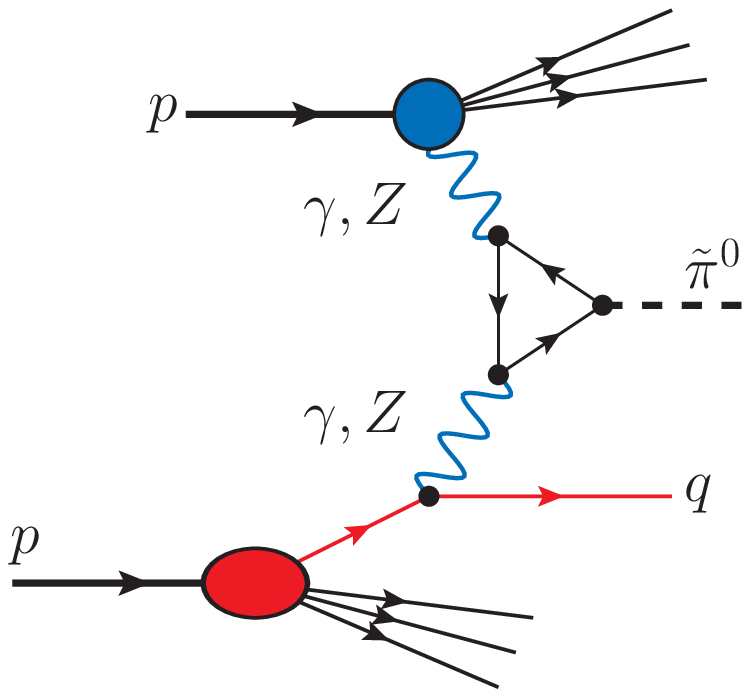}
\includegraphics[width=0.2\textwidth]{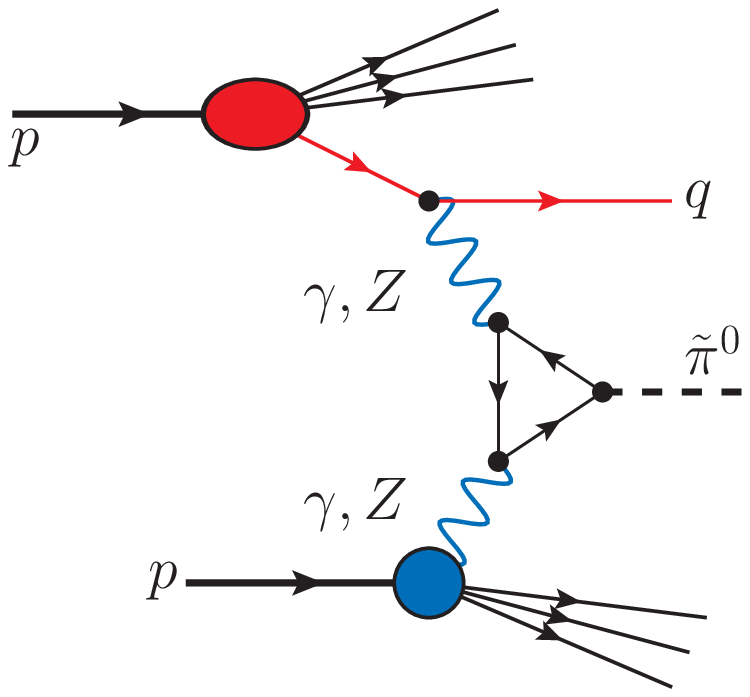}
  \caption{\label{fig:diagrams_order2}
  \small
Technipion production via the $2 \to 2$ partonic subprocesses.}
\end{figure}

\begin{figure}[!ht]
\centering
\includegraphics[width=0.2\textwidth]{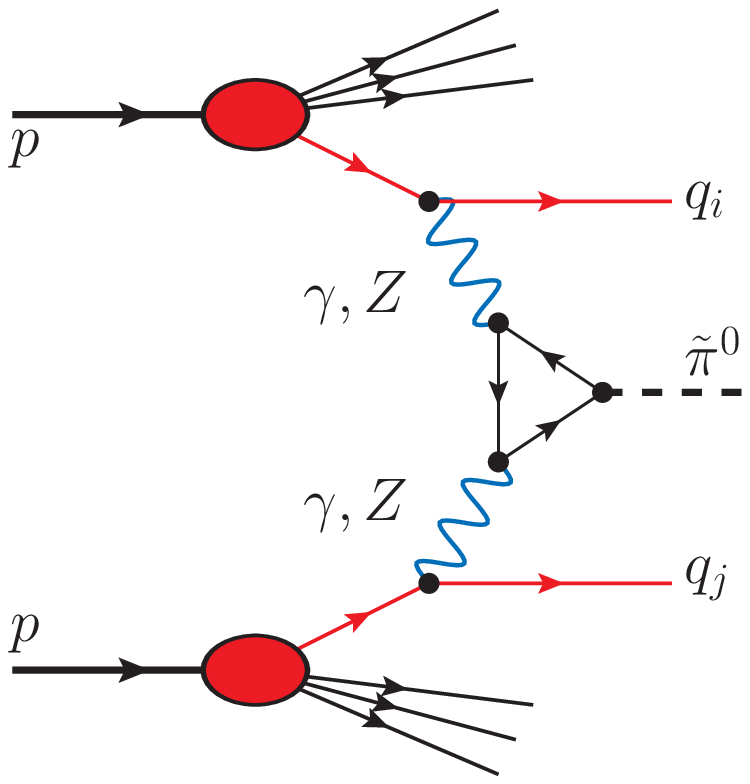}
  \caption{\label{fig:diagrams_order3}
  \small
Technipion production via the $2 \to 3$ partonic subprocesses.}
\end{figure}

The mechanisms considered in our recent paper \cite{Lebiedowicz:2016lmn} are shown in 
Figs.~\ref{fig:diagrams_order1}-\ref{fig:diagrams_order3}.

\section{An example of the amplitude calculation}

In the case of VTC technipion model \cite{Pasechnik:2013bxa}, 
the amplitude for the $\gamma \gamma \to {\tilde \pi}^0 \to \gamma \gamma$ subprocess reads:
\begin{eqnarray}
&&{\cal M}_{\gamma\gamma \to {\tilde \pi}^0 \to \gamma\gamma}
(\lambda_1, \lambda_2, \lambda_3, \lambda_4) =
(\varepsilon^{(\gamma)\mu_{3}}(p_{3},\lambda_{3}))^{*}  
(\varepsilon^{(\gamma)\mu_{4}}(p_{4},\lambda_{4}))^{*}\,
\nonumber\\
&&
\times 
\epsilon_{\mu_{3}\mu_{4} \nu_{3}\nu_{4}} p_{3}^{\nu_{3}} p_{4}^{\nu_{4}}\,
F_{\gamma\gamma}\, 
\frac{i}{\hat{s}-m_{\tilde{\pi}^{0}}^{2}+i m_{\tilde{\pi}^{0}}\Gamma_{tot}} \;
\epsilon_{\mu_{1}\mu_{2} \nu_{1}\nu_{2}} p_{1}^{\nu_{1}} p_{2}^{\nu_{2}}\,
F_{\gamma\gamma}\,
\varepsilon^{(\gamma)\mu_{1}}(p_{1},\lambda_{1})
\varepsilon^{(\gamma)\mu_{2}}(p_{2},\lambda_{2})\,,
\label{amp_signal}
\end{eqnarray}
where the effective neutral technipion coupling $F_{\gamma \gamma}$ is \cite{Pasechnik:2013bxa}
\begin{eqnarray}
F_{\gamma\gamma}=
\frac{4\alpha_{em}\,g_{\rm TC}}{\pi}\frac{m_{\tilde{Q}}}{m_{\tilde{\pi}^{0}}^2}\,
\arcsin^2\Bigl(\frac{m_{\tilde{\pi}^{0}}}{2m_{\tilde{Q}}}\Bigr)\,, \qquad
\frac{m_{\tilde{\pi}^{0}}}{2m_{\tilde{Q}}}<1\,.
\label{F_gamgam}
\end{eqnarray}

The $\Gamma_{tot}$ can be calculated from a model or taken from recent experimental data. 
In the following we take the calculated value of $\Gamma_{tot}$ and $m_{\tilde{\pi}^{0}}$ = 750 GeV.
The mass scale of the degenerate techniquarks $m_{\tilde Q}$ 
is in principle another free parameter (see Ref.~\cite{Lebiedowicz:2013fta}).

The cross section for the signal 
is calculated as ($\mu_F^2 = p_{t,\gamma}^2$):
\begin{equation}
\frac{d \sigma}{d y_3 d y_4 d^2 p_{t,\gamma}}
= \frac{1}{16 \pi^2 {\hat s}^2} \sum_{ij}
x_1 \gamma^{(i)}(x_1,\mu_F^2) x_2 \gamma^{(j)}(x_2,\mu_F^2)
\overline{|
{\cal M}_{\gamma\gamma \to \tilde{\pi}^{0} \to \gamma\gamma} |^2} \,,
\label{inclusive_gamgam_signal}
\end{equation}
where $i, j = $ el or in, i.e. they correspond to elastic or inelastic fluxes
($x$-distributions) of equivalent photons, respectively, and
$x_{1}$, $x_{2}$ are the longitudinal momentum fractions of the proton
\begin{eqnarray}
x_1 = \frac{p_{t,\gamma}}{\sqrt{s}} \left[ \exp({y_{3}}) + \exp(y_{4}) \right]\,, \quad
x_2 = \frac{p_{t,\gamma}}{\sqrt{s}} \left[ \exp({-y_{3}}) + \exp(-y_{4}) \right]\,.
\end{eqnarray}
%

\section{Selected results}
\label{section:Results}

We summarize our results in Table~1
where we have collected cross sections for different QED orders shown in the figures above.
The elastic photon fluxes were calculated using the 
Drees-Zeppenfeld parametrization \cite{Drees:1988pp,Drees:1994zx}, 
where a simple parametrization of the nucleon electromagnetic form factors is used.
To calculate inelastic contributions we use collinear approach 
with photon MRST(QED) parton distributions \cite{Martin:2004dh}. 
Surprisingly, different contributions are of the same order of magnitude. 
In this calculation $g_{TC} = 10$ and $m_{{\tilde Q}} = 0.75 \,m_{{\tilde \pi}^{0}}$ were used. 
To describe the experimental signal more precisely $g_{TC}$ can be rescaled. 

\begin{table}
\begin{small}
\textbf{Table~1}: Hadronic cross section in fb for
technipion production for different contributions, see Figs.~\ref{fig:diagrams_order1}-\ref{fig:diagrams_order3}.
\end{small}
\begin{footnotesize}\begin{center}
\begin{tabular}{|l|c|c|c|c|c|}
\hline
Component                           &$\sqrt{s} = 1.96$ TeV &$7$ TeV &$8$ TeV  &$13$ TeV &$100$ TeV   \\
\hline
2 $\to$ 1 (in, in)                & 1.37$\times 10^{-3}$        & 0.16 & 0.22    & 0.55 &  8.08       \\
2 $\to$ 1 (in, el)                & 0.22$\times 10^{-3}$        & 0.05 & 0.06   & 0.15 &  1.88       \\
2 $\to$ 1 (el, in)                & 0.22$\times 10^{-3}$        & 0.05 & 0.06     & 0.15 &  1.88       \\
2 $\to$ 1 (el, el)                & 0.03$\times 10^{-3}$        & 0.01 & 0.02     & 0.04 &  0.42       \\
\textbf{2 $\to$ 1,  sum of all}  & \textbf{1.84$\times$ 10$^{-3}$} & \textbf{0.27} & \textbf{0.36}     & \textbf{0.89} &     \textbf{12.26}       \\
\hline
2 $\to$ 2 (in, in), two diagrams         & 0.74$\times 10^{-3}$ & 0.14 & 0.19 & 0.49 & 7.69  \\
2 $\to$ 2 (in, el) and (el, in)          & 0.13$\times 10^{-3}$ & 0.05 & 0.07 & 0.19 & 2.93  \\
\textbf{2 $\to$ 2,  sum of all}          & \textbf{0.87$\times$ 10$^{-3}$} & \textbf{0.19} & \textbf{0.26}  & \textbf{0.68} &    \textbf{10.62}  \\
2 $\to$ 2,  sum of all, $p_{t,jet} >$ 10 GeV&         &&       & 0.43 & 8.03       \\
2 $\to$ 2,  sum of all, $p_{t,jet} >$ 20 GeV&         &&       & 0.35 & 6.99       \\
2 $\to$ 2,  sum of all, $p_{t,jet} >$ 50 GeV&         &&       & 0.25 & 5.42       \\
\hline
\textbf{2 $\to$ 3}                           &   \textbf{0.14$\times$ 10$^{-3}$}      & \textbf{0.09}& \textbf{0.13}      & \textbf{0.46}  &  \textbf{16.71}   \\
2 $\to$ 3,  $p_{t,jet} >$ 10 GeV             &         &&       & 0.04  &  1.41     \\
\hline
\end{tabular}
\label{table:table1}
\end{center}
\end{footnotesize}
\end{table}

\begin{figure}[!ht]
\centering
\includegraphics[width=0.42\textwidth]{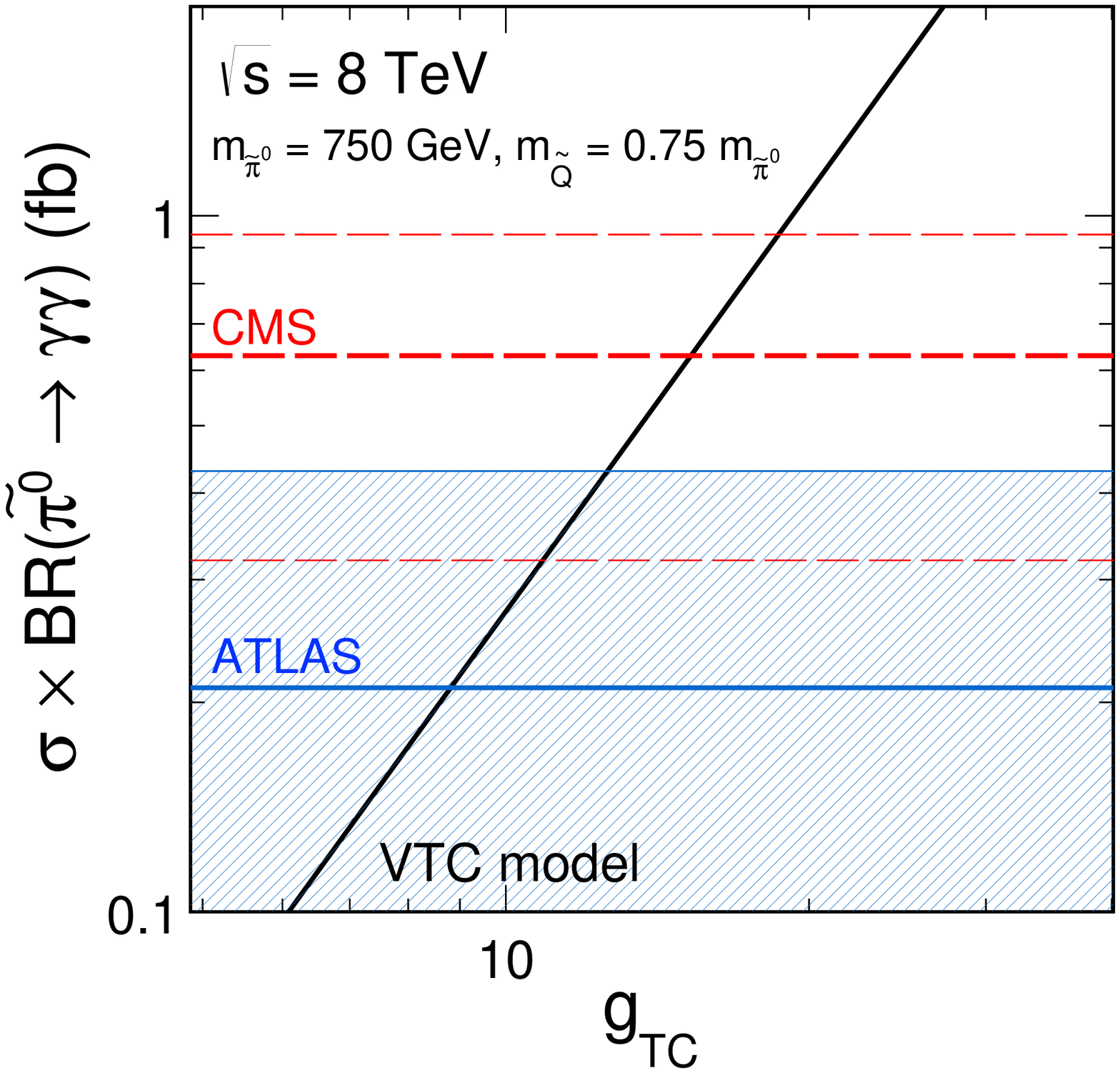}
\includegraphics[width=0.42\textwidth]{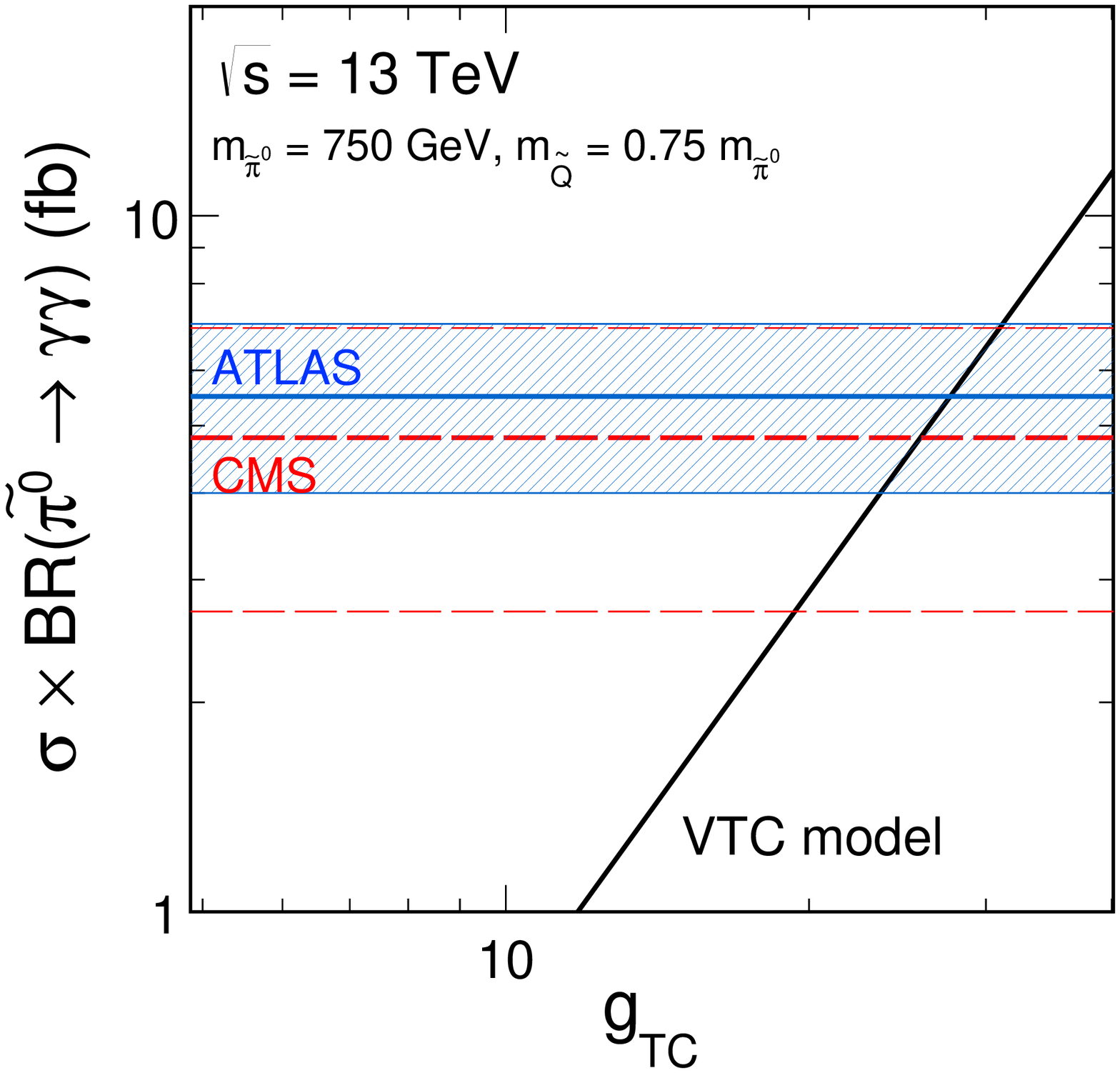}
  \caption{\label{fig:hadronic_g_TC}
  \small
The dependence of the hadronic $p p \to ({\tilde \pi}^0 \to \gamma \gamma) + X$ cross section 
on $g_{TC}$ together with the crudely estimated in \cite{Strumia:2016wys} 
experimental result at the LHC \cite{Aaboud:2016tru, Khachatryan:2016hje}. 
The solid black line represents our result for the technipion production in the VTC model.
}
\end{figure}

The dependence of the cross section on $g_{TC}$ is shown in Fig~\ref{fig:hadronic_g_TC} 
for $\sqrt{s} = 8$~TeV (left panel) and $\sqrt{s} = 13$~TeV (right panel)
within an experimental uncertainties taken from \cite{Strumia:2016wys} (narrow width scenario).
Our result for $g_{TC}$ = 20 and our standard choice $m_{\tilde Q} = 0.75\, m_{\tilde \pi}^{0}$
is at the lower edge of experimental uncertainties
at $\sqrt{s} = 13$~TeV and at the upper edge of experimental uncertainties at $\sqrt{s} = 8$~TeV.
If $m_{\tilde Q}/m_{\tilde \pi}^{0}$ is smaller the $g_{TC}$ could be lower,
see Fig.~8 of \cite{Lebiedowicz:2013fta}.
The value of $g_{TC}$ = 20 could be smaller when exchange of $Z$ bosons is included. 

\begin{figure}[!ht]
\centering
\includegraphics[width=0.4\textwidth]{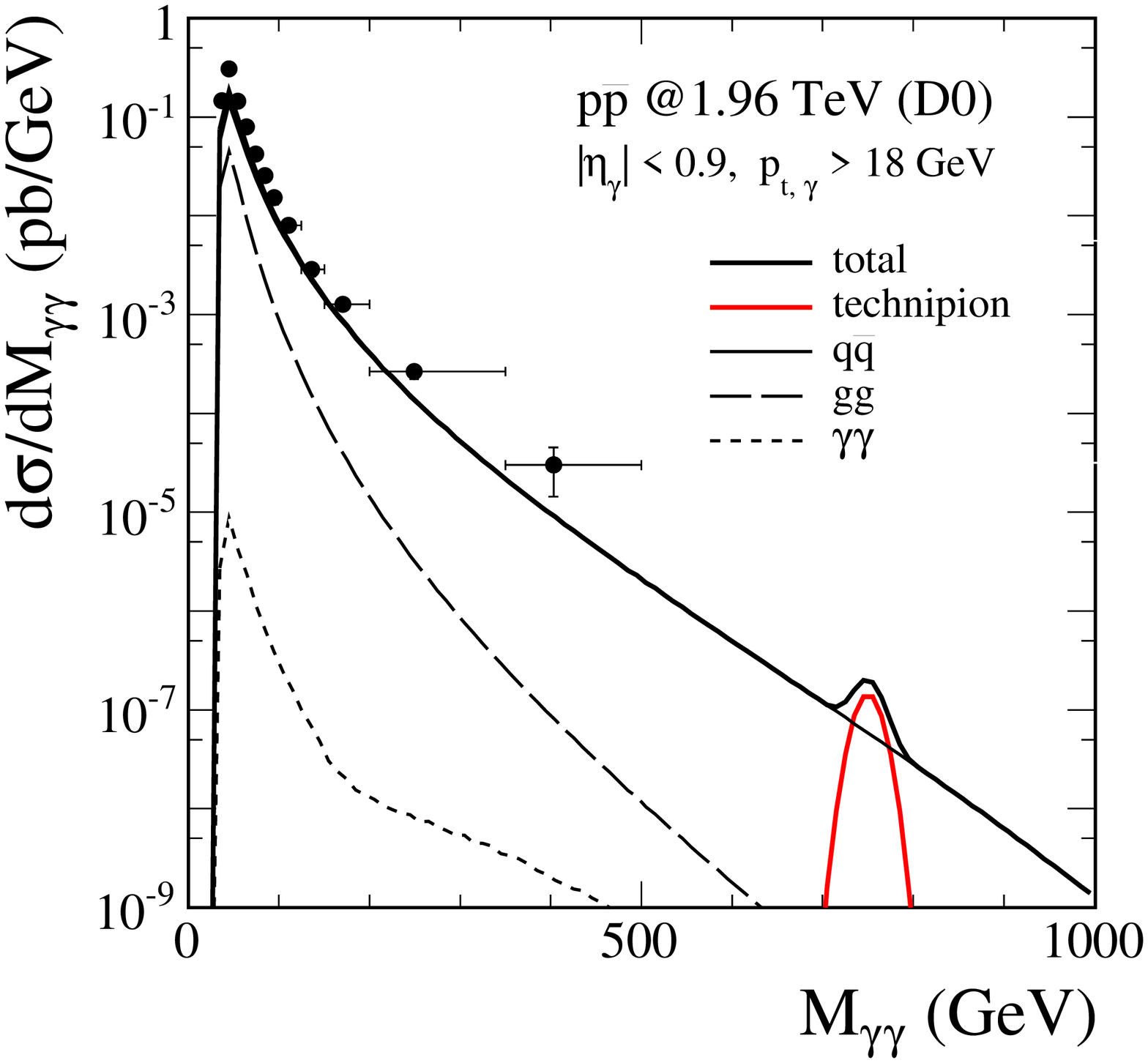}
\includegraphics[width=0.4\textwidth]{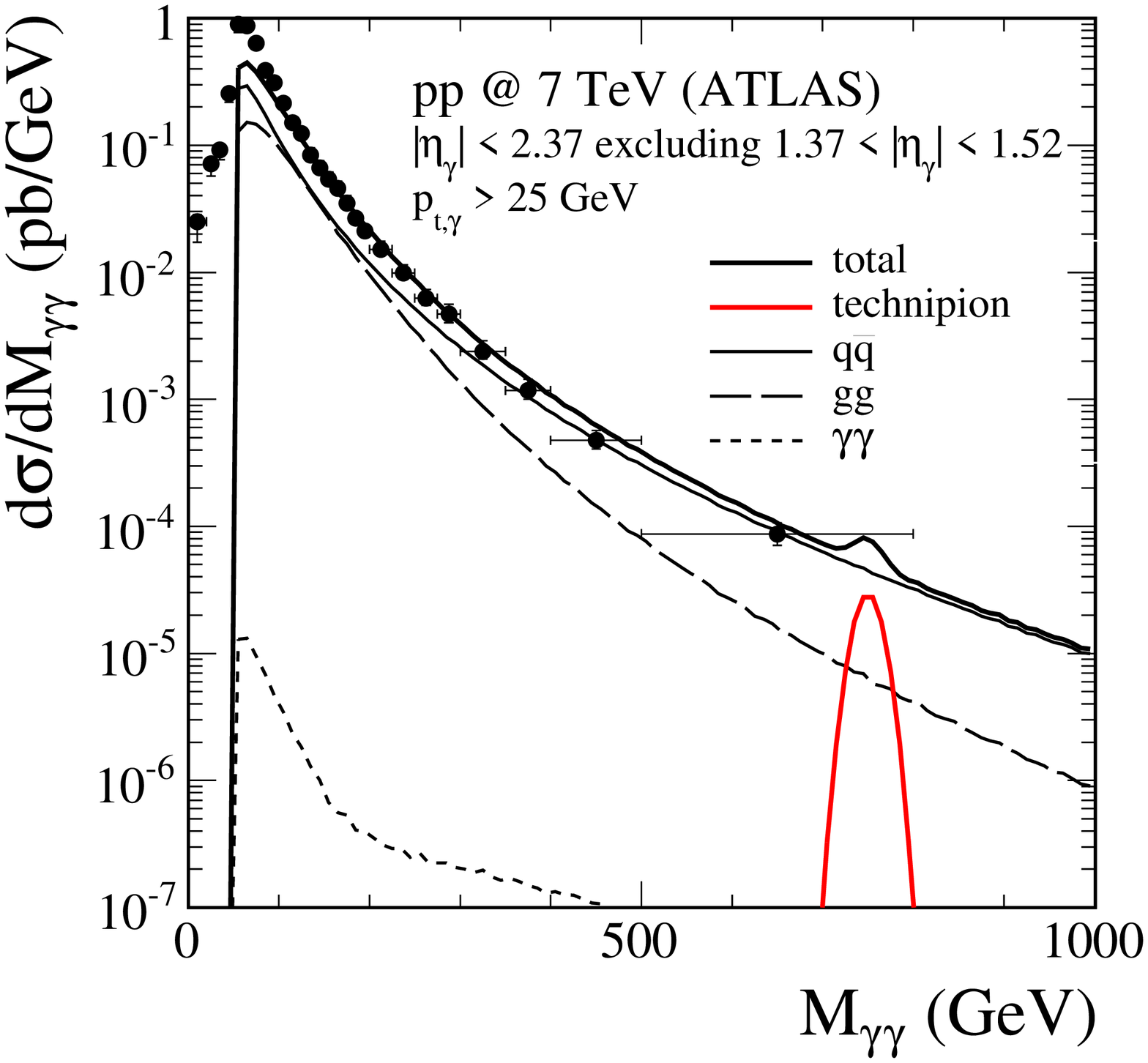}\\
\includegraphics[width=0.4\textwidth]{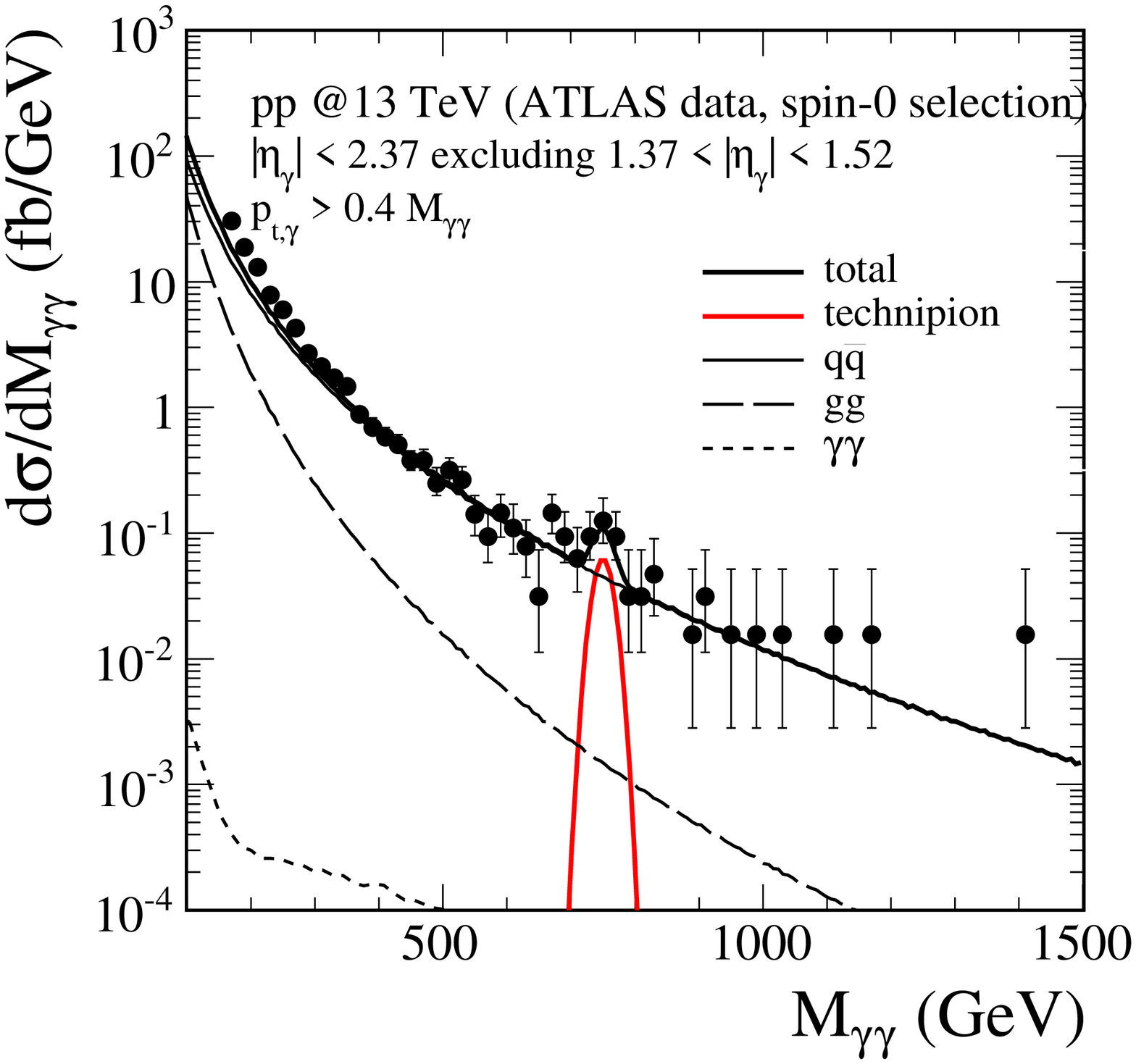}
\includegraphics[width=0.4\textwidth]{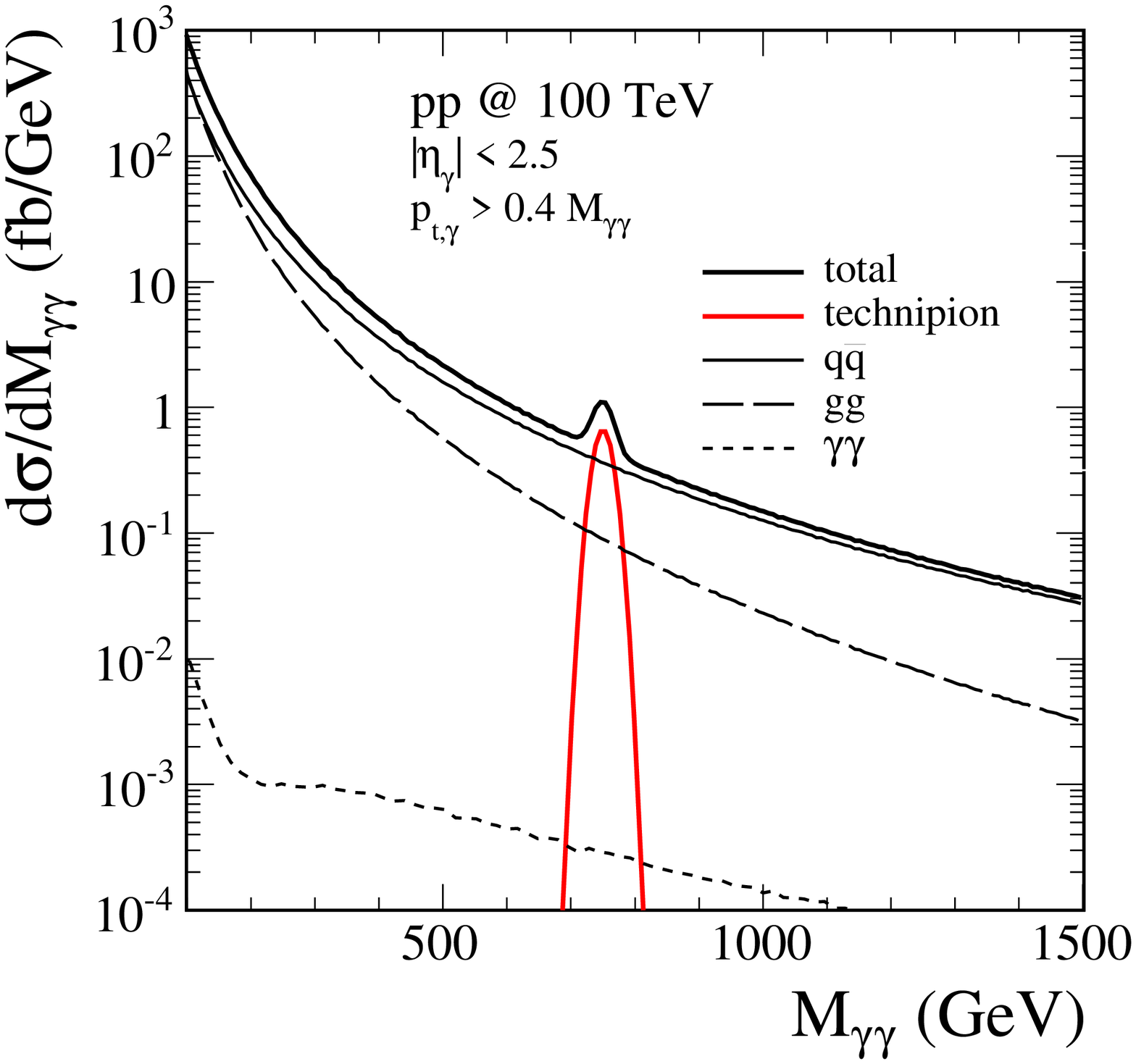}
  \caption{\label{fig:M_gamgam}
  \small
The two-photon invariant mass distributions 
for different background contributions
and the signal-technipion predictions obtained in the VTC model
including experimental cuts.
For comparison, the experimental data from D0 \cite{Abazov:2013pua} at $\sqrt{s}$ = 1.96 TeV,
ATLAS at $\sqrt{s}$ = 7 TeV \cite{Aad:2012tba}, 
the recent ATLAS data (spin-0 selection) at $\sqrt{s}$ = 13 TeV \cite{Aaboud:2016tru}
and our prediction for Future Circular Collider are presented.
}
\end{figure}

The most important is the distribution in diphoton invariant mass
where the signal was observed.
In Fig.~\ref{fig:M_gamgam} we show four examples relevant 
for different experiments using their kinematic conditions:
D0 at $\sqrt{s}$ = 1.96 TeV \cite{Abazov:2013pua},
ATLAS at $\sqrt{s}$ = 7 TeV \cite{Aad:2012tba},
ATLAS at $\sqrt{s}$ = 13 TeV \cite{Aaboud:2016tru},
and the prediction for Future Circular Collider at $\sqrt{s}$ = 100 TeV.
We show both signal and background contributions.
Clearly the $q \bar q$ annihilation contribution dominates, especially
at large invariant masses in the surrounding of the signal. 
In our analysis the experimental invariant mass resolution was included for the signal-technipion calculations
in the following simple way
\begin{eqnarray}
\frac{d \sigma}{dM_{\gamma\gamma}} 
= \sigma_{\tilde{\pi}^{0}} \, \frac{1}{\sqrt{2 \pi} \sigma} \exp\left({\frac{-(M_{\gamma\gamma} - m_{\tilde{\pi}^{0}})^{2}}{2 \sigma^{2}}}\right)\,.
\label{resolution}
\end{eqnarray}
In the calculation we take $\sigma = 15$~GeV assuming $\sigma/m_{\tilde{\pi}^{0}} \sim$ 2\%.
In Eq.~(\ref{resolution}) we take $\sigma_{\tilde{\pi}^{0}}$ = 0.005~fb, 1.09~fb, 2.36~fb, 24.83~fb
corresponding to $\sqrt{s}$ = 1.96, 7, 13, 100 TeV, respectively,
including the relevant kinematical cuts shown in the panels of Fig.~\ref{fig:M_gamgam}.
The values of cross sections above were obtained from Eq.~(\ref{inclusive_gamgam_signal})
and $g_{TC} = 20$.

\section{Conclusions}

In our recent paper \cite{Lebiedowicz:2016lmn} 
we discussed a possibility that recently
observed by the ATLAS and CMS Collaborations diphoton enhancement 
at invariant mass $M_{\gamma \gamma} \approx$ 750 GeV is a technipion. 
The main emphasis was put on chirally-symmetric (vector-like) Technicolor model with 
two mass degenerate (techni)flavours. In this model only 
$\gamma \gamma$, $\gamma Z$ and $Z Z$ couplings and production
mechanisms are possible. Therefore the decay width is rather small 
$\Gamma_{tot} \ll$ 1 GeV.

We discussed there in detail the production mechanisms within the considered model. 
In the present version we included only photon initiated processes.
In some modern parton distribution models photons are included as partons
in the proton. In this model there is a reach pattern of electroweak
contributions. We have considered $2 \to 1$, $2 \to 2$ and $2 \to 3$ type
processes. We have found that they give similar contributions
to the cross section. In order to describe the observed ``signal''
we had to adjust model coupling of techniquarks to the neutral 
technipion. Including the photon initiated processes only we have found
that $g_{TC} \simeq$ 20 is not inconsistent with the experimental data.


In addition, we have made predictions for the Tevatron, Run-I LHC 
and Future Circular Collider. The predictions for the Tevatron have been
discussed in the context of existing data in the diphoton channel.
We have concluded that the cross section for energies lower than 8 TeV
are so small (below background for integrated luminosity limit) 
that the signal could not be observed. 

We have also made predictions for purely exclusive case.
We have predicted the cross section of the order of 0.2~fb at $\sqrt{s} = 13$~TeV.
To focus on such a case one has to measure technipion (two photons) 
in the central detectors as well as both protons in forward directions. 

In Ref.~\cite{Lebiedowicz:2016lmn} we considered also an alternative one-family walking 
technicolor model discussed recently in Ref.~\cite{Matsuzaki:2015che} (see also references therein). 
In this model the gluon-gluon fusion is the dominant 
production mechanism of assumed isoscalar technipion. 
We refer to \cite{Lebiedowicz:2016lmn} for details of the corresponding analysis.

In summary, the considered technicolor models cannot be excluded
by the present $\gamma \gamma$ and dijet data \cite{Aaltonen:2008dn, Aad:2013tea}.

This research was partially supported by the Polish National Science
Centre Grants DEC-2013/09/D/ST2/03724 and DEC-2014/15/B/ST2/02528 and by the Centre 
for Innovation and Transfer of Natural Sciences and Engineering Knowledge in Rzesz{\'o}w. 
R. P. was partially supported by the Swedish Research Council Grant No. 2013-4287.

\end{document}